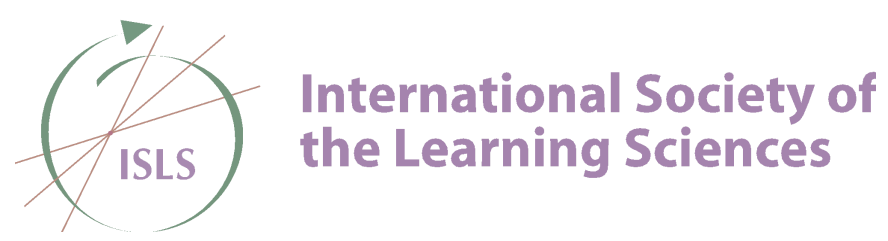

# "You Can Actually Do Something": Shifts in High School Computer Science Teachers' Conceptions of AI/ML Systems and Algorithmic Justice

Daniel J. Noh, University of Pennsylvania, dnoh@upenn.edu
Deborah A. Fields, Utah State University, deborah.fields@usu.edu
Yasmin B. Kafai, University of Pennsylvania, kafai@upenn.edu
Danaé Metaxa, University of Pennsylvania, metaxa@upenn.edu

**Abstract:** The recent proliferation of artificial intelligence and machine learning (AI/ML) systems highlights the need for all people to develop effective competencies to interact with and examine AI/ML systems. We study shifts in five experienced high school CS teachers' understanding of AI/ML systems after one year of participatory design, where they co-developed lessons on AI auditing, a systematic method to query AI/ML systems. Drawing on individual and group interviews, we found that teachers' perspectives became more situated, grounding their understanding in everyday contexts; more critical, reflecting growing awareness of harms; and more agentic, highlighting possibilities for action. Further, across all three perspectives, teachers consistently framed algorithmic justice through their role as educators, situating their concerns within their school communities. In the discussion, we consider the ways teachers' perspectives shifted, how AI auditing can shape these shifts, and the implications of these findings on AI literacy for both teachers and students.

## Introduction

The recent proliferation of artificial intelligence and machine learning (AI/ML) systems in so many aspects of people's professional, public, and private lives highlights the need for all to develop effective competencies to interact with and examine AI/ML systems (Long & Magerko, 2022), a goal that has been espoused around the globe (UNESCO, 2023). One emphasis in the growing conceptualization of AI literacy is that people need to understand how AI/ML is used in various applications and industries and how it influences their lives and society at large (Touretzky et al., 2019). Over the past five years, many initiatives have focused on supporting *students'* AI literacy, particularly through tool development (e.g., Teachable Machines: Carney et al., 2020) and curricular design (e.g., Touretzsky, 2022; DAILy Curriculum, 2025; Druga et al., 2022), alongside a steadily growing recognition of the need to foster a more critical AI literacy (e.g., McClure et al., 2025; Solyst et al., 2025; Veldhuis et al., 2025). Critical AI literacy is key to computational empowerment, helping to build awareness of the "ethical consequences" in the designs and uses of AI/ML technologies, including algorithmic injustice, where harmful bias can affect both society and everyday life (Iivari et al., 2025). Algorithmic injustice arises when social and political judgments are encoded into algorithmic systems under the guise of neutrality, reproducing historical inequalities and power asymmetries (Birhane, 2021). Yet, while efforts to define and cultivate AI literacy for K-12 learners have begun expanding, far less attention has been paid to supporting and understanding computer science (CS) *teachers'* critical AI literacy development.

CS teachers are vital to fostering students' critical engagement with AI/ML systems, yet little research has attended to how CS teachers develop their own understanding and engagement with these systems. Recent research has demonstrated that CS teachers in the U.S. are aware of the importance of AI but "less than half feel equipped to teach it" (Maslej, 2025, p. 378). Only a few studies have focused on supporting teachers' technical understandings of AI/ML systems (e.g., Chounta et al., 2022; Sanusi et al., 2021; Velander et al., 2024) or on promoting trust and use of AI/ML in classrooms (e.g., Kizilcec, 2024; Williams et al., 2022). Furthermore, based on a 2022 survey of CS teachers across the U.S., less than half cover digital citizenship, ethics of computing, or impacts of computing in their classes (Koshy et al., 2022), consistent with prior landscape surveys that show a majority of U.S.-based CS teachers generally have not seen the importance of covering computing's role in perpetuating biases and other inequities in the classroom (Koshy et al., 2021).

In this paper, we study shifts in five experienced CS teachers' understanding of AI/ML systems over one year of participatory design, during which they co-developed lessons on AI auditing as a systematic method to query AI/ML systems that affect youth (see next section). Following Kafai and Proctor's (2021) call for "alternate endpoints of what it means to be computationally literate," being inclusive of learners' "identities and their respective communities," we take a broad stance toward teachers' critical AI literacy. Namely, we recognize that, as AI/ML systems impact both society and daily life, critical AI literacy must similarly span across teachers' professional, public, and private lives, and be situated within their school communities' cultures and political priorities. Drawing on pre- and post-interviews as well as a collective, member-checking focus group interview, we invited teachers to reflect on their experiences with algorithmic (in)justice, AI/ML systems, and classroom practices. Specifically, we ask: (1) How did CS teachers develop their critical understanding of and engagement with AI/ML systems over the course of a year? (2) How were their conceptions situated in their professional, public, and private lives as well as within their school communities? Through our analysis, we found that teachers shifted towards more situated, critical, and agentic perspectives of AI/ML systems, contributing insights into how AI auditing can foster critical AI literacy among teachers.

## Background

Current research suggests that many teachers possess limited understanding of issues such as algorithmic bias, algorithmic injustice, and techno-solutionism (Hu & Yadav, 2023) and feel discomfort with integrating ethics in AI/ML lessons, fearing they might say something incorrect (Kizilcec, 2024). Our approach to studying CS teachers' critical AI literacy was centered on two key concepts—algorithmic justice as "relational ethics" (Birhane, 2021), which grounds "ethics, justice, knowledge, bias, and fairness in context, history, and an engaging epistemology" (p. 2) and "computational empowerment" which focuses on developing skills and reflexivity to understand and engage "critically, curiously, and constructively" with digital technology in everyday life and society at large (Smith et al., 2023). AI/ML technologies have specifically been called out for their potentially harmful algorithmic biases, automating and perpetuating historical, unjust and discriminatory patterns (Birhane, 2021, p. 1). In a relational ethics perspective, rethinking ethics with AI/ML systems requires zooming out to examine foundational questions, unstated assumptions, asymmetrical power dynamics, and ingrained social and structural inequalities (p. 8), rather than "devising ways to 'debias' datasets or derive abstract 'fairness' metrics." Thus, fostering critical AI literacy, where practicing reflection enables learners to "analyze, question, critique, and transform the information they encounter" (Veldhius et al., 2024, drawing on Freire, 1970), becomes necessary. In studying teachers' conceptions of algorithmic justice from the framing of relational ethics, then, it is important to consider their whole lives, beyond curricular content and classroom practice, situating AI/ML systems and algorithmic justice across teachers' everyday experiences (Hawkins & Pea, 1987).

We also draw on the framework of computational empowerment, foregrounding its emphasis on cultivating skills and reflexivity by engaging with both the construction and deconstruction of technologies (Smith et al., 2023). In the context of AI/ML, AI auditing (also called algorithm auditing) is a professional method that, when extended to include everyday users, can foster computational empowerment's focus on critical examination or deconstruction (Morales-Navarro et al., 2024). Expert-led AI audits involve the systematic and empirical evaluation of algorithmic systems to "draw conclusions about the algorithm's opaque inner workings and possible external impact" (Metaxa et al., 2021). Notably, audits are conducted without direct system access, usually by independent third parties. Experts have used audits to identify biases or illegal discrimination in AI/ML systems ranging from face detection tools (Buolamwini & Gebru, 2018) to online ad delivery platforms (Ali et al., 2019).

While AI auditing has largely remained a method for experts, recent audits have included everyday people, enabling them to conduct structured evaluations of AI systems modeled after expert audits (Lam et al., 2022). End-users possess "rich situated knowledge of the particular impacts that algorithmic systems have on their own communities" that expert auditors may not be aware of (Lam et al., 2022, p. 512:2). Many harmful AI/ML system behaviors are difficult to detect outside of the specific situated contexts of users, as demonstrated by a number of "everyday audits" by users, including Twitter cropping algorithms, Yelp algorithms, recommendation systems, and banking or credit scores (Shen et al., 2021). Shen and colleagues (2021) describe such auditing practices as forms of "everyday algorithmic resistance." Building on De Certeau (1984) and Scott's (1985) accounts of "everyday resistance," such practices involve subtle acts of questioning and pushing back against dominant power structures in the mundane contexts of daily life, rather than, say, organized opposition to AI/ML systems. In this sense, end-user and everyday audits form a unique intersection between a method developed by expert auditors and the knowledge of everyday users.

Building on this growing research, over the past few years emerging research has shown the potential for youth to engage in and contribute to AI auditing (Morales-Navarro et al., 2025a; Morales-Navarro et al., 2025b; Solyst et al., 2025b). Drawing on participatory design with teens, Morales-Navarro and colleagues (2025a) developed five steps of auditing that can help scaffold youth participation: developing a hypothesis, generating a set of systematic inputs, running a test, analyzing the outputs, and reporting the results. They found that "incorporating aspects of expert-driven auditing" while also "drawing from user-driven auditing practices" supported youth in conducting original audits, building on youths' everyday experiences with AI/ML systems

(p. 29192). Similarly, Solyst and colleagues (2025b) found that “youth can contribute quality insights” shaped by their specialized knowledge, lived experiences, and age-related knowledge (p. 2098). Thus, AI auditing has demonstratively supported learners’ computational empowerment through critiquing and assessing AI/ML technologies they use in everyday life. In this paper, we examine how teachers—through learning about AI auditing, designing lessons, and implementing them in their classrooms—can apply their everyday and professional knowledge to engage in everyday algorithmic resistance, and how these practices contribute to their development of critical understanding and computational empowerment over the course of a year.

## Methods

### Positionality statement

We consider teachers as whole people with experiences that weave in and out of their roles as teachers, applying this perspective in our study design and analysis. Our long relationships with the teachers, stretching 6-11 years over multiple collaborations, helped build the trust required to invite reflections that extend across their lives. We seek to maintain these relationships and prioritize their perspectives and problems of practice. Thus, during the AI auditing participatory design work, we foregrounded teachers’ ideas and curricular interests in design at every stage, acknowledging our own power and status in academia. Teachers were paid at a competitive hourly rate for all time contributed to the project outside their teaching jobs.

We also recognize that our own identities and backgrounds shape our research approach. Our team represents at least five racial/ethnic identities, three gender identities, and academic expertise in the learning sciences and human-computer interaction (HCI). Most of our team resides in the Mid-Atlantic region of the United States, within a few hours of two of the teachers, while the second author lives in the same West Coast metropolitan region within an hour of the three remaining teachers and has been in their classrooms repeatedly over a decade. Our qualifications—including experience conducting expert audits, teaching high school students, and designing learning environments—enabled us to carry out this study responsibly and effectively.

### Context and participants

This study was conducted within a larger participatory design project on AI auditing, grounded in principles that foreground teacher and youth voices in computing education (Smith et al., 2023), with the goal of creating lessons for high school CS classrooms that foster critical engagement with AI/ML systems.

For this work, we recruited five public school teachers who had prior experience teaching equity-driven physical computing in introductory computer science classes (e.g., Fields et al., 2018). Teachers brought substantial experience across a range of contexts (two areas of the US; spanning from metropolitan to rural) and brought diverse perspectives on the design and implementation of AI auditing activities. Listed below are anonymized descriptions of participating teachers, including characteristics of the schools they teach at (e.g., racial demographics in descending order; * signifies majority low-income students):

- **Becca**: Asian woman, 17 years teaching (9 CS)
  - School: metropolitan*, US West Coast (Latinx, Black, Multiracial)
- **James:** Multiracial man, 22 years teaching (12 CS)
  - School: metropolitan*, US West Coast (Latinx, Black, White, Polynesian)
- **Ben:** Japanese & Mexican man, 15 years teaching (12 CS)
  - School: metropolitan*, US West Coast (White, Latinx, Black, Asian, Multiracial)
- **Alyssa:** Black & Cherokee woman, 6 years teaching (6 CS)
  - School: suburban, US Mid-Atlantic (White, Asian, Black, Latinx, Multiracial)
- **Chloe:** White & Greek American woman, 14 years teaching (4 CS)
  - School: rural*, US Mid-Atlantic (White, Black, Latinx, Biracial)

Two teachers began the study with greater knowledge and confidence in AI/ML. James had experience in industry prior to his teaching career and had many close friends in CS, engineering, and science with whom he regularly discussed issues of ethics in AI/ML, especially regarding environmental and human labor concerns. Ben had a masters in CS from a prominent university and had taken a semester-long graduate course on AI. The other three teachers generally expressed having less experience with AI/ML but were eager to learn.

The lesson design work included several phases. We began with a weeklong, in-person professional development workshop (Summer 2024) that included opportunities to engage in playful AI auditing activities, hear from experts on AI auditing (faculty and graduate students), learn from youth-advisors who had engaged in AI auditing during earlier informal workshops (Noh et al., 2025), and participatory design sessions where teachers and researchers brainstormed, outlined, and crafted a set of four classroom lessons. Design continued into the fall with virtual meetings until lessons were fully drafted. From Winter 2024 to Spring 2025, all teachers implemented the lessons in their own classrooms. This was followed by a second weeklong, remote professional development workshop (Summer 2025), which again included meeting experts, iterative design and reflection sessions, and a final focus group interview (see below).

## Data and analysis

Data collection took place in three phases, in the form of interviews about the teachers' understanding of algorithmic justice as well as their experiences with AI/ML systems and/or algorithmic justice in both their professional (teaching) contexts and personal lives. Phase 1 included individual, hour-long semi-structured pre-interviews on Zoom before the first professional development workshop (June-July 2024). After initial analysis and work with the teachers over one year, for Phase 2 we adapted the interview protocol, repeating some questions and adding new ones to follow up on some initial themes from early analysis (e.g., the situatedness of teachers' perspectives of algorithmic justice) and to provide an opportunity for teachers to share how their ideas about algorithmic justice and AI/ML systems had shifted. Phase 2 post-interviews were again an hour-long and over Zoom, scheduled immediately after each teacher completed implementation of AI auditing lessons in their classrooms (April-June 2025). After initial analysis of post-interviews, for Phase 3 we developed a focus group interview protocol sharing initial themes and insights alongside questions to facilitate a conversation where teachers could collectively speak to and build on (or refute) these initial insights (July 2025). Of note, all but one teacher (Becca) participated in the final focus group. All interviews were led by the first and/or second authors, transcribed by Zoom, then manually edited for accuracy.

Our analysis draws on iterative, thematic qualitative analysis (Braun & Clarke, 2006; Braun et al., 2022) involving two rounds of emergent coding between the phases of data collection and a further three formal rounds of analysis after data collection was complete. While the first and second author led analysis, they consulted with the broader research team at each stage, sharing analysis, examples, and insights throughout the process. After editing the transcripts and conducting preliminary analysis after the final interview (focus groups, Phase 3), we identified four emergent themes related to teachers' perceptions of algorithmic justice and AI/ML systems, particularly around the ideas of situatedness, criticality, and agency, as well as a special characteristic of teachers' perspective related to their role as teachers. We then conducted a first round of coding (1) looking at the ways in which these four themes occurred in the data and (2) inspired by Lee et al.'s (2021) humanistic stance toward data science education, examining whether teachers' reflections were articulated at varying scales: a personal layer (everyday experiences), a cultural layer (teaching practice, community), or a sociopolitical layer (societal discourse, broader media).

At this stage, we recognized that in order to track changes over time we needed to compare each individual teacher's pre- and post-interviews, so in a second round of analysis, we comparatively examined teachers' pre- and post- interviews individually, tracking changes across examples of AI/ML systems they referenced and perceptions of algorithmic justice they exhibited. Exemplary excerpts were selected to illustrate these shifts and to highlight how teachers' understandings were situated within the varying personal, cultural, and sociopolitical layers. The third round of analysis involved focused, thematic coding across the dataset. We reorganized and refined earlier codes into a systematic scheme, allowing us to explicate four overarching themes, capturing how teachers understood and engaged with algorithmic justice in their practice and everyday experiences. To ensure analytical consistency, two researchers separately coded one teacher's pre- and post-interviews and reconciled minor discrepancies through discussion and refinement of the codebook before the first author applied the scheme across the full dataset.

# Findings

We describe the teachers' shifts in understanding of algorithmic justice and AI/ML systems after a year of engaging with the design and teaching of lessons on AI auditing. Analysis of pre- and post-interviews revealed three emerging patterns: teachers' perspectives became (1) more situated, connecting their understandings to everyday experiences and contexts; (2) more critical, showing a deeper awareness of harms in AI/ML systems; and (3) more agentic, recognizing possibilities for action and change. Further, across all three patterns, (4) teachers consistently framed algorithmic justice through their professional role as educators, with attention to their responsibility to prepare students to recognize and respond to harmful algorithmic systems. In the following sections, we first detail how their perspectives became more situated, more critical, and more agentic, and then highlight how their framing of algorithmic justice was further shaped by their role as teachers.

## From broad ideas to personal examples: Situating algorithmic justice in everyday experiences

One significant change in the teachers' understanding emerged from the examples the teachers provided about algorithmic injustice, shifting from more broad understandings of AI/ML systems informed by public narratives to more situated understandings based on their everyday, personal experiences with these systems. In early Summer 2024, during pre-interviews, the teachers demonstrated a relatively strong awareness of societal consequences of harmful bias, but these were largely "big picture" examples of harmful AI/ML systems that were separate from their personal everyday experiences, including predictive policing, financial loan approval, teacher and student evaluation metrics, and racial discrimination in healthcare. Several teachers expressed how these examples came from news and popular media, such as prominent expert voices (Joy Buolamwini), books (Weapons of Math Destruction by mathematician Cathy O'Neill), and television shows (Grey's Anatomy). Even when teachers drew on their own observations, their responses remained broad and speculative. For instance, James theorized that college admission algorithms might be affecting his students, but did not share any more direct encounters with AI/ML systems. In contrast, in Summer 2025, teachers focused conversations around examples from their everyday experiences with such systems, including discussions about using generative AI to plan family trips, encounters with deep fakes on social media, and even experiences with adjusted prices when purchasing car insurance.

To illustrate this shift, we share examples from Chloe. In her pre-interview, when Chloe was asked to describe any harmful biases or discrimination from AI/ML systems, she mentioned "high end" examples, including misidentification in facial recognition technologies at stores and airports and the use of fallible technologies in criminal justice, where people are "accused of doing something that they didn't do based on… [waves her hands] data." While this example demonstrates her knowledge of harmful bias in AI/ML systems, her gesturing conveyed her feeling of a lack of expertise about what specific data might be used in this system and where it came from. She mentioned a few personal examples, such as the irrelevance of autocomplete suggestions on her mobile device, but these were minor and had more to do with the technical limitations of AI/ML systems than potential harm. In contrast, her examples in the post-interviews consisted mainly of directly experienced, everyday AI/ML systems found across her car, social media feeds, targeted advertisements, streaming services, her school's student data-tracking system, concert ticket sales systems, and grocery stores. Notably, these examples came, not from other people's reports, but from her own personal observation and analysis. Within these everyday AI/ML systems, she questioned the kinds of data being collected by ambient systems with which she interacted (i.e., car, social media, advertisement, grocery store) and shared concerns with how these AI/ML systems may induce harm to herself, her family or her students (i.e., data privacy for students, pricing disparity for concert tickets, discomfort with certain targeted ads).

This shift reflects a move from understanding harms from AI/ML systems through distant authority to developing situated, evidence-based insights grounded in the teachers' own everyday interactions. By critically engaging with personal experiences, teachers connected algorithmic justice to the context of their own lives and communities. While addressing large-scale changes that deal with unjust surveillance or harmful biases in facial recognition technologies is, understandably, a daunting task, addressing the more personal, immediate injustices paves the path toward agency and activism—a theme discussed in more depth later in the Findings.

## From unsuspecting to critical awareness: Critically engaging with AI/ML systems

Another shift was a growth in teachers' critical awareness of AI/ML systems and algorithmic harm. Here, critical awareness represents a reflective understanding of how these systems operate, including their limitations and potential for harm, coupled with the ability to question the impact of such systems rather than simply accepting them as they are. In pre-interviews, several teachers emphasized how little they knew about AI/ML systems, describing themselves as "starting at a pretty blank slate" (Becca) or admitting that their "knowledge, when we started, was zero" (Chloe). They framed their relationship to algorithmic injustice with hesitation, claiming, "I'm sure that I've experienced it and may not have realized it" (Alyssa). Notably, these experienced computer science teachers expressed conscious awareness of their lack of knowledge about these systems. However, in the post-interviews, the teachers shifted to actively interrogating the fairness and implications of these systems.

Consider Becca, who, over the course of the year, began to question outputs from AI/ML systems consistently across her family life and teaching practice. When reflecting on her prior perceptions of AI/ML systems, she noted how, with these systems, she didn't "really, like, ask questions" about AI or "dive a little deeper into" it. She further explained that, by the end of the year, she regularly scrutinized AI tools rather than take their outputs at face value. In her personal life, she recounted her husband using an AI-based itinerary generator to plan an overseas family vacation. While she acknowledged that the tool synthesized useful information, she argued with her husband that the AI tool could never capture her "human knowledge," including how much walking she preferred or how she likes to travel. "My feelings are not in the sources," she explained, highlighting the limitations of reliance on AI/ML systems. Similarly, Becca sought to instill a more questioning attitude with her students. Noting that her students regularly used Google Gemini, as it automatically appeared at the top of Google search queries, she prodded her students to evaluate the often misleading or incomplete information: "Whatever AI tool I'm using. I don't take it as a fact." Becca's growing critical awareness led her to approach AI/ML systems with skepticism rather than reject them, while also encouraging her family and students to think more critically about their outputs.

Even one teacher who started the study with a fairly high level of knowledge about AI/ML systems (with a master's degree in computer science, which included a graduate course on AI ethics) expressed growth in critical understanding. James explained how he began viewing ordinary image search results with a more critical lens. "Sometimes you see the results, and you're just like, well, okay, sure. 'Skateboarding'—a bunch of young kids." But looking more closely, he might notice patterns that could be potentially harmful, pointing to how a search result of professional skateboarders may show mostly male or light-skinned individuals. He connected this realization to questions of representation, wondering, "If you really don't look like this, you may feel discouraged to go into this career." What had once seemed like neutral outputs became evidence of how even search engines could reinforce societal biases, prompting a deeper awareness to recognize subtle but potentially consequential harms in everyday AI/ML systems. These examples mark progression from unsuspecting encounters with AI/ML systems and justice to more deliberate forms of critical awareness, stretching across teachers' professional, public, and private lives (Kafai & Proctor, 2021). At the same time, this burgeoning criticality accompanied a growing sense of hope.

## From helplessness to action: Finding hope through agency

In addition to increased critical awareness of AI/ML systems, each teacher experienced profound shifts from helplessness to hope and agency with algorithmic systems. This sense of agency is substantiated in both their descriptions of potential action towards algorithmic (in)justice and their budding confidence in engaging with conversations with others about algorithmic (in)justice.

In pre-interviews, all teachers expressed some form of powerlessness in the face of AI/ML systems. Even James and Ben, who had the most prior knowledge about AI/ML, felt helpless. James mentioned with concern that he had not "really heard a lot of examples of how they fixed any algorithms." He stated that "it would be great if it wasn't just reducing bias. But if it was eliminating it, or helping people that are [...] being harmed." Similarly, when Ben was asked to share any examples "where algorithms worked for people or were equitable or accountable," he defaulted to examples of algorithmic injustice, rather than justice, concerning issues of excessive human labor from AI/ML model training and intellectual property infringement. In their pre-interviews, both teachers depicted a sophisticated understanding of "relational ethics" (Birhane, 2021), yet felt powerless to act. By Summer 2025, all teachers expressed a stronger sense of agency with AI/ML systems. For instance, James expressed that, due to his newfound knowledge of AI auditing, he felt "less helpless" and that "people actually can [...] make some change." Similarly, Ben stated that his best takeaway was that, even with the ubiquity of these systems, at least "we have some way of maybe putting a little bit of a curb on it." AI auditing provided a means to "do something" (James) about harmful biases in AI/ML systems.

One specific area of action teachers mentioned was developing the confidence to speak with others about algorithmic injustice. For Chloe, this meant the self-assurance to name and point to examples of harmful algorithmic systems at school: "I feel like it's easier to talk about." Ben similarly highlighted a shift in his teaching, noting that "now that I have more tools and more ways of talking about it than I did before," asking students pertinent questions "so that maybe they'll think a little bit more." Alyssa explained that this new assurance in having discussions about AI/ML emerged from new knowledge gained during lesson design when experts talked about how systems like resume screening actually work. Having someone "walk through it" not only gave her deeper understanding but also a way to communicate that understanding with others. Moving from helplessness to agency, the teachers developed a new capacity to articulate concerns and engage in dialogue about algorithmic justice in their everyday and school communities.

## Algorithmic justice framed by their role as teachers

Across their reflections, the teachers consistently framed their understanding of algorithmic justice through their professional roles as educators, specifically attuned to their school communities' cultures and needs. Becca and Ben, for instance, brought forward concerns from the very beginning of the project that were specific to the metropolitan, majority Black and Latinx student population of their schools, including harmful biases in racial profiling in policing. Becca noted that her students "had a lot of high emotion when it comes to the police and profiling" and questioned, "how can you go to someone and say, 'Hey, you're about to commit a crime [when] they haven't committed any crime yet." Similarly, in describing algorithmic injustice, Ben noted everyday examples that his students might face, such as walking into a store and being followed because someone suspects they might shoplift, noting how AI/ML systems amplify these biases. A related motivation expressed by several teachers was understanding issues of AI/ML bias in specific careers. For instance, Alyssa and Becca both worked in schools with specific emphases on biomedical fields, and they sought to learn about AI/ML systems in these fields to support their students' career goals.

Further, the teachers expressed their desire to help students question and reflect on the ways AI/ML systems can shape young people's everyday lives. Indeed, teaching about AI/ML became a space to invite students into critical conversations about harms derived from AI/ML systems. Chloe, for instance, described how her experiences learning and designing lessons about AI auditing gave her confidence to more comfortably "have those conversations with students" and to ask, "Okay, well, is this correct? What would you do in this situation?" In this, her role as a teacher helped turn her growing critical understanding of AI/ML systems into actionable classroom dialogue, supporting students in questioning and reflecting on potential harms.

Finally, teachers framed their role as enabling students to imagine and agentically enact change. James explained his responsibility as a CS teacher "to make sure people are aware and realize they may be able to take some kind of action." In learning about AI audits, both formal and informal, he saw a "path to activism," highlighting that people, including himself and his students, can provide feedback to companies to push for change, declaring that, "hey, you can actually do something." Alyssa similarly emphasized that it was important "as a teacher, being able to teach students about this and the power that they have and the difference they can make when it comes to algorithms." In these and similar statements, AI auditing became a framework for empowering students to recognize inequities and take meaningful action. All of these examples show how teachers positioned themselves as both learners of algorithmic justice *and* as educators, foregrounding their students' needs and contexts in their thinking.

## Discussion

In this paper, we examined high school computer science teachers' shifts in their understanding of algorithmic justice before and after they became engaged in AI auditing activities. We found that teachers' perspectives became more *situated*, framed by their understanding in everyday experiences and contexts; more *critical*, reflecting growing awareness of harms in AI/ML systems; and more *agentic*, highlighting possibilities for action and change. In all three perspectives, teachers consistently framed algorithmic justice through their professional role as educators, situating their concerns within their students' and school communities. In the following, we discuss teachers' ways of knowing and reasoning about AI/ML systems shifted, the role of AI auditing in such shifts, and the larger implications of AI literacy for teachers and students.

Our most important goal was to develop a better understanding of what high school CS teachers know and think about algorithmic justice—their understanding of how algorithmic systems could operate in fair and accountable ways but also produce harm by considering historical injustices and power asymmetries in computing. While the issue of inequities has been part of a longstanding discussion and efforts in broadening participation in computing (Margolis et al., 2008), how this applies to AI/ML systems has not followed suit. Many of the teachers' initial understandings reflected general perceptions as they relate to issues derived from media discourse (e.g., talks by Joy Buolamwini). However, through closer engagements with AI auditing, their understandings became increasingly evidence-based and contextually-grounded in their own and their students' experiences. One interpretation of this shift could be that teachers moved from relying on authority accounts to using evidence provided by multiple experiences and observations of their own and of their students' interactions with AI/ML systems. In this case, teachers' growing confidence came from realizing that they could analyze, question, and interpret AI/ML systems independently, rather than relying solely on the experiences and words of experts. These shifts also reflect engagement with relational ethics (Birhane, 2021), as teachers began to understand algorithmic justice in terms of relationships, contexts, and their responsibilities to their students and communities—centering these relational dimensions over technical faults of AI/ML systems when discussing algorithmic justice.

The teachers' shift towards critical questioning and speculations about AI/ML systems demonstrated a form of everyday algorithmic resistance. To review, everyday algorithmic resistance refers to subtle, often unnoticed ways people question or push back against dominant algorithmic systems in their everyday contexts (Shen et al., 2021). While conducting AI audits might reveal algorithmic bias or even harmful outputs, such activities do not provide a direct means to address a system's shortcomings in the way that, for example, retraining a model with a revised dataset might. Yet, through these engagements, teachers began to recognize avenues for agency. That is, although auditing alone does not provide a means to "fix" algorithms, it opens up opportunities for teachers to act and have agency within their professional roles. By engaging students in reflecting on systems, ranging from image searches to surveillance technologies, teachers found that teaching about AI auditing may help both them and their students push against potential harms from AI/ML systems and the ways these systems shape everyday life. This sense of agency was uniquely tied to their position as educators, as they found power through action—specifically, through teaching and designing lessons. Additionally, engaging teachers with designing lessons about AI audits introduced them to a new approach used in human-computer interaction studies but not in computer science education. By positioning teachers as co-designers, the PD project may have enabled them to adapt expert auditing practices for their own classrooms, selecting examples and questions that resonated with their students and local contexts. Although such analysis was beyond the scope of this study, future research could examine how the participatory design process itself contributed to these shifts in agency and criticality.

Finally, we address what critical AI literacy for teachers can mean. Because teachers act as key mediators of students' AI literacy, these shifts in teachers' perspectives are meaningful in shaping how students develop their understanding and engagement with AI/ML concepts in the classroom. By integrating their situated, critical, and agentic understandings into their teaching, teachers may create richer opportunities for students to engage with these perspectives in their own learning. Thus, future work should examine how such shifts in teacher perceptions translate into classroom practice and how they shape students' development of situated, critical, and agentic conceptions of algorithmic justice. Further, research across contexts, including geographic regions, different sized classrooms, and varying grade bands, could help us understand how different settings support or limit such processes. Finally, understanding which aspects of the professional development setup—whether it was the participatory design, expert engagement, or collaborative efforts—helped foster these shifts will be essential for future designs of effective and supportive activities for and with teachers.

## Conclusion

In this paper, we examined how experienced computer science high school teachers' understanding of algorithmic justice shifted after designing and implementing classroom lessons on AI auditing. We found that their conceptions of algorithmic justice over time became more situated, personal, and included more student-relevant examples and experiences. These shifts were supported by teachers' active engagement with AI auditing, which provided opportunities to both critically analyze and interrogate AI/ML systems themselves and to guide their students in doing the same. Through this process, teachers gained a sense of agency in navigating and addressing harms from AI/ML systems, translating ethical considerations into actionable classroom practices. The examples provided in the paper illustrate that learning about algorithmic justice can foster computational empowerment, a key dimension of critical AI literacy, enabling teachers to question and act on the societal impacts of AI/ML systems.

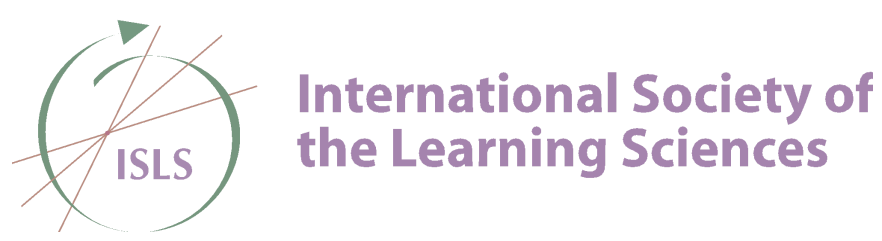

## Acknowledgements

Initial data collection (pre-interviews) and participatory design in 2024 were supported by an NSF grant to Yasmin Kafai and Danaé Metaxa (#2342438). Later data collection and analysis were supported by Rapid Response Bridge Funding from the Spencer Foundation, the Kapor Center, the William T. Grant Foundation, and the Alfred P. Sloan Foundation. Any opinions, findings, and conclusions or recommendations expressed in this paper are those of the authors and do not necessarily reflect the views of NSF, the Kapor Center, William T. Grant, Spencer, or Alfred P. Sloan Foundations, the University of Pennsylvania, or Utah State University. Special thanks to Luis Morales-Navarro and the Learning Sciences RAC at Penn for feedback on earlier versions of the paper. Most of all, our deepest appreciation to the five teachers who engaged in design, implementation, and this research over the course of a year.